\documentclass[12pt,a4paper]{article}
\usepackage{jheppub}

\usepackage{graphicx}
\usepackage{subcaption}
\usepackage{xspace}
\usepackage[titletoc]{appendix}

\usepackage[utf8]{inputenc}

\usepackage[style=numeric-comp,sorting=none]{biblatex}
\abstract{The rapid evolution of technology and the parallel increasing
complexity of algorithmic analysis in HEP requires developers
to acquire a much larger portfolio of programming skills. Young
researchers graduating from universities worldwide currently do not receive adequate
preparation in the very diverse fields of modern computing to respond to growing needs
of the most advanced experimental challenges. There is a growing consensus in
the HEP community on the need for training programmes to bring researchers up to date with new software technologies,
in particular in the domains of concurrent programming and artificial intelligence. We review some
of the initiatives under way for introducing new training programmes and highlight some of the issues
that need to be taken into account for these to be successful.}

\begin{document}

\begin{tabular*}{\linewidth}{lc@{\extracolsep{\fill}}r@{\extracolsep{0pt}}}
 & & HSF-CWP-2017-02 \\
 & & June 20, 2018 \\ 
 & & \\
\end{tabular*}
\vspace{2.0cm}

\title{HEP Software Foundation Community White Paper Working Group --
Training, Staffing and Careers}


\author[1]{Dario Berzano,}
\author[2]{Riccardo Maria Bianchi,}
\author[3]{Peter Elmer,}
\author[4]{Sergei V. Gleyzer}
\author[1]{John Harvey,}
\author[5]{Roger Jones,}
\author[6]{Michel Jouvin,}
\author[7]{Daniel S. Katz,}
\author[8]{Sudhir Malik,}
\author[9]{Dario Menasce,}
\author[7]{Mark Neubauer,}
\author[10]{Fernanda Psihas,}
\author[11,a]{Albert Puig Navarro,}
\author[1]{Graeme A. Stewart,}
\author[12]{Christopher Tunnell,}
\author[10]{Justin A. Vasel,}
\author[4]{and Sean-Jiun Wang}

\affiliation[1]{CERN, Geneva, Switzerland}
\affiliation[2]{Department of Physics and Astronomy, University of Pittsburgh, Pittsburgh, PA, USA}
\affiliation[3]{Princeton University, Princeton, NJ, USA}
\affiliation[4]{University of Florida, Gainesville, FL, USA}
\affiliation[5]{Lancaster University, Lancaster, UK}
\affiliation[6]{LAL, Université Paris-Sud and CNRS/IN2P3, Orsay, France}
\affiliation[7]{University of Illinois Urbana-Champaign, Urbana, IL, USA}
\affiliation[8]{University of Puerto Rico, Mayaguez, PR, USA}
\affiliation[9]{INFN Sezione di Milano-Bicocca, Italy}
\affiliation[10]{Department of Physics, Indiana University, Bloomington, IN,
USA}
\affiliation[11]{Physik-Institut, Universität Zürich, Zürich, Switzerland}
\affiliation[12]{University of Chicago, Chicago, IL, USA}
\affiliation[a]{Supported by SNF under contract 168169.}

\maketitle

\clearpage

\section{Introduction}

The field of high energy physics (HEP) has consistently exploited
the latest innovations in computational tools and technologies for processing data.
The software toolset of the particle physicist is ever-growing, and the problem
set increasingly complex. Young researchers will likely encounter a number of
programming techniques with which they are totally unfamiliar, and will therefore
need ``on-the-job'' training in order for them to be productive.

Evidently a strong research community comprises independent thinkers who ask
research questions and direct their own analyses, instead of blindly following
prescriptions and recipes. It is therefore imperative that training opportunities
and resources are made available, such that adequate career
path possibilities exist for people in HEP who would otherwise leave the field
due to limited advancement opportunities.

To produce the best possible science, it is important that physicists can easily acquire
essential software engineering skills. The nature and level of the expertise a
physicist needs to acquire will vary according to the problems that need to be tackled.
For example, basic data analysis requires familiarity with the Python programming language
and with a broad range of data analysis libraries in current usage. However, those
wishing to contribute to large open source projects, such as ROOT, require greater
training in the use of software engineering practices that cover the full development
life-cycle. Clearly a thorough understanding of the physics problem domain is
essential, making it necessary to provide a career path for those
young physicists willing to invest time and effort in becoming software specialists.

In each HSF workshop, a consensus on the need for proper tutoring, training,
and relevant career possibilities emerged as the basic ingredients for ensuring success.
The main requirements and challenges to be faced were identified as follows:

\begin{itemize}
    \item to encourage and provide incentives to the members of the HEP
    community to train students and other collaborators,
    \item to properly assign credit to software development as a scientific
    discipline, including training activities,
    \item to establish policies for the hiring and long-term retention of researchers
    specialising in computing,
    \item to address the gap in formal software training that is not always
    given by universities as part of a physicist's education.
\end{itemize}

\section{Career Paths for Software Experts}


The introduction and development of sophisticated software tools has led to
improvements in the performance of our data processing software. The
personnel who drive these advances through novel implementations work on a broad
range of tasks, including the development of high speed DAQ systems, of modern databases for
storing the state of a detector, the management of data flow from the first trigger
level to the final analysis dataset, as well as the development of algorithms and
mathematical tools for extracting publishable physics results.

As tools become more complex, physicists must be continuously retrained in order
to utilize them effectively. It is important to find a way to deploy training efforts
holistically and cross-experimentally, whenever possible, in order to avoid unnecessary
duplication of the effort involved. This will require organization and coordination to
ensure that different software teams can agree on common standards, thereby allowing different
experiments to adopt the same solutions in an efficient way.

Beyond the immediate benefit to the scientific community in having well-trained
collaborators, most people who start a physics graduate programme will have
careers outside of academia. This trend is reflected in several calls for
proposals from national scientific funding agencies that promote training in computer
infrastructure areas, such as the EU's Horizon 2020 programme and those from
the National Science Foundation (NSF) \cite{NSF1,NSF2}.

A key goal is to find incentives that will encourage
increasing numbers of people to dedicate their time and effort to train their
colleagues.
To be effective, the training must be done by people who work at the cutting
edge of the technology and who in many cases are found among the youngest
researchers. These are also the very same people who are most in need of
officially recognized credits for the advancement of their careers. Currently,
visibility and recognition is given mainly to those working on data analysis
projects, rather than to the development and support of the underlying
software. One solution to this problem would be the establishment of specific
career paths for researchers specialising in scientific software development.
However, the practical
implementation of such an approach is extremely challenging, as there are a wide range of
institutions involved that belong to different countries with their own
policies, priorities, and funding strategies.

Broadly speaking, two scientific profiles of researchers could be
envisaged. A physicst could make a detailed plan of what they need, expressed
in a tidy requirements document, and a computer scientist could use this document
to provide the required solution. This is how the software development process
typically works in the software industry. However, in HEP requirements are rapidly
changing, forcing developers and physicsts to interact closely during the entire
process of software development. This works smoothly only when both communities
have the same goals and speak the same language, hence the need for physicists with a good
knowledge of computer science.

The first possible career path would be that of a \emph{physicist with computing
science specialisation}, also known as a physicist-programmer in some
communities. (It would be conceptually different to that of a \emph{computer
scientist} as the two have different goals, seek different paths to the solution
of their problems, and usually do not even share the same language.) Such people
would have an active role in physics analysis, and so meet many of the criteria
needed for an academic career path.

The second would be the path of a domain-specific \emph{software
engineer}. Such a person has the primary role of developing software and finding
software solutions, but with a large domain-specific knowledge to
understand relevant use cases and the available solutions.
Such posts are almost impossible to establish in current academic
institutions; rather being seen by many funding agencies and
universities as short term technical positions. This does not allow domain
specific expertise to be aquired or retained in our field.

Both career paths suffer in academic terms from few or poorly cited
publications, few opportunities to win grants in their own right and little
opportunity for impact with commerce and wider society. These three failings
make an academic appointment challenging or disfavoured as compared to other
applicants. However, there are measures that can be taken to address all three
weaknesses, which we expand on in the remainder of this document.

\section{Training needs of the community}

The HEP community consists of people with diverse software
experience, interests, and time availablity to learn new techniques.  Any
training programme must take into account the variation in target audiences. For
example, an undergraduate doing a summer research project has different needs
and skills than their professor.

\subsection{Classification of trainees}

For purposes of training we can broadly classify four
different experience levels:

\begin{itemize}
   \item \emph{Beginner}: New collaborators with no knowledge of the
   tools or techniques they are expected to use. These are people in need of some kind of formal
   training in modern computing techniques, such as compiled and scripting
   languages, together with operating system basics such as filesystems, version
   control, and command-line shells.
   \item \emph{Intermediate}: People with some experience in
   concepts and tools, but looking to supplement their
   experience with more recent and modern approaches.
   \item \emph{Advanced}: Experts who have mastered current
   technologies and implementations and who want to stay up-to-date with
   new advanced developments.
   \item \emph{Software Specialists}: HEP scientists in charge of software
   development in areas not limited only to analysis, such as DAQ systems,
   computing infrastructure, databases, pattern recongition, and complete
   frameworks.
\end{itemize}

Each of these groups has very different training needs. However,
\emph{whenever possible}, any training programme should take advantage of
developments in pedagogy, such as active
learning~\cite{ActiveLearning}, peer learning~\cite{PeerLearning}, and
web-based training.\footnote{See \S\ref{sec:initiatives}.} In some cases, it may
even be advantageous to hand out code samples that are purposely broken or
flawed, and ask students to fix or improve them. Learning the material in a way
that sticks is difficult and challenging for both the students
and the instructor and often takes more time than we would prefer. However, this
is the best way to educate scientists who can fully
contribute to the physics programmes at large, which is really the ultimate goal
of any training programme.

\subsection{Knowledge that needs to be transferred}

At all stages of software and computing training, we should take care to
encourage \emph{good practices across the community}, such as ensuring error
checking, modularity of code design, version control, writing tests, etc. All
the key concepts addressed in training should not be specific to a particular
experiment or field of application, but general enough to be useful to the whole
HEP community and beyond. A number of specific concepts need to be taught, in
order to guarantee the basic level of competence needed to write efficient code
for the various tasks that need to be performed in HEP experiments.
These include
programming concepts, data structures, basics of code design, error checking,
code management tools, validation and debugging tools. More advanced topics
include modularity of code design, advanced data structures, evaluation metrics,
writing tests and working with different types of hardware accelerator.
Additionally, special emphasis should be given to
reporting results and documenting them.

Some of the training subjects considered important to pursue are
listed in Appendix \ref{appendix-a---training-topics-of-interest}.

\section{Advancing Training in HEP}

The implementation of training should employ different training formats, such as
videos, wikis, lectures, jupyter notebooks and advanced visualizations, etc.,
so that people can learn in a familiar and effective manner and in such a way
that experts are encouraged to share their knowledge.

An important point to consider is the difficulty, often experienced in the past,
in developing large software programs across different experiment
collaborations.
While there already are experiment-specific training efforts in place, there are
many needs that are in common. Establishing a common training programme could
help facilitate the sharing of exprience amongst different experiments. This
would reduce duplication of efforts and enable growth of a shared training
culture by accumulating and sharing expertise.
To realise this goal, a possibility could be the creation of a
\emph{federation} of training initiatives,
aimed at improving the efficiency and cost-effectivness of this important
activity. An appropriate incentive programme to reward those who train the community
might help to facilitate the goals of such an initiative.

When trying to exchange training materials within a group the first problem is
to convince authors to contribute their work. This issue
can be alleviated by addressing the correct assignment of intellectual property.
Another problem is that trainers do not like to reuse given material as-is. They usually
want to refactor it, building their own training history.
This makes it difficult to have everyone agree on common approaches.
It is even challenging to agree to host material in the same
centralized place. A way to overcome this could be to settle for a centralized
catalog, with each author being free to host their material in their location
of choice.

Another challenge is the wide range of student competence. Special care must be
given to setup a training structure that can manage both introductory and
advanced material. This may be addressed by organising course material as a
large number of independant topics. The IN2P3 authors try to restrict their
tutorials to 25 minutes, as inspired by the Pomodoro
technique\cite{PomodoroTechnique}, so that one can easily jump and compose one's
own curriculum. However, it turns out to be tricky to keep such small tutorials
really independant and meaningful when they can be selected at will.

All trainers have also faced the tremendous loss of time
in software installation for any practical exercises as many students have
not managed to prepare their machines in advance.
Here containers may bring real
progress, provided that everyone has at least Docker\cite{Docker} installed on
their machine.
JupyterHub\cite{JupyterHub} is a technology which supports training sessions
without the need for specialist installations.

\subsection{Initiatives for Future Training Programmes\label{sec:initiatives}}

Some methods that can be used for location independent training include Massive
Online Open Courses (MOOCs), hands-on workshops, online knowledge bases, expert trainer volunteer
networks, and web-based training approaches.

MOOCs can be used to develop an open-source set of
tutorials and tools. Existing online courses such as Udacity~\cite{Udacity} and
Coursera~\cite{Coursera} can be
evaluated and exploited by the community. A novel approach such as
WikiToLearn~\cite{WikiToLearn} could also be explored to assess potential benefits (as has already been attempted
in the context of the GridKa School of Computing~\cite{GridKa}, see below).

Experiment-specific and global knowledge bases can be established with
incentives for experts to contribute. They can be open-source so that a lot of
knowledge can be added by the trainees themselves as information is learned;
this is the particular context where an approach such as WikiToLearn could be
of great help.

Question-and-answer websites such as Stack Overflow~\cite{Stackoverflow} for HEP are also a very
useful resource for common problems and questions. This approach has already
been considered by HSF start-up members, and turned out to be difficult to
pursue due to the lack of a critical mass needed by Stack Overflow, but in the
future, boundary conditions might change, making this approach viable.

Hands-on workshops are an invaluable part of learning how to apply theoretical
concepts in practice. Identifying which of the current workshops are productive
and useful, and if they cover all the topics that need to be transferred and
that are in demand by the students, is an important action item.

Creating an expert tutor volunteer network is another way to provide training
and support to the community.
This of course requires proper recognition, at least in terms of career
prospects as an incentive for young researchers. It is clear that the best
possible tutors are, in principle, those people who are actively engaged in
modern software developments: these are often young researchers in the
first steps of their careers and, in order to be attracted to devoting
substantial time to training and tutoring, they must be assured that such an
effort would be properly recognised in an official way. Such kind of recognition
is not currently implemented, at least not in a standardized or official way. A
possible structure for such a network could be the establishment of a
{\em federation} of existing schools, as discussed in \S\ref{sec:ehcurtrainprog}.

\subsection{Web Based Training}

Difficulties that have emerged in the past with respect to implementing training
courses are the lack of funding and the lack of available time by experts in the
field. People with enough expertise or insight usually don't have time to devote
to prolonged periods of student training, and, even when they can find time, the
cost of setting up a training course in an effective way is often beyond what is
made available by funding agencies (funds for travel, hosting, setting up a room
with a computing infrastructure to allow interactive hands-on session, etc.) A
possible solution is a completely different approach to training, using a
web-based platform to provide training materials to students. This would be
complementary to the already existing and successful efforts such as the CERN
School of Computing's Bertinoro and KIT ones.

The web based approach has several advantages over traditional ones:
\begin{itemize}
   \item Tutors can add material to the web site at a very slow pace (whenever
   they find time to do it, one slide a week or a chapter a day) .
   \item Their material, publicly available on a web-site, can be further
   expanded by collaborators (also at their own pace) or, even better and more productively, by students
   who decide to contribute new additions, examples, exercises etc. Such a
   collaborative effort allows more people to be exposed to training at any
   given time, creates a sense of community, and creates bridges between people
   in contiguous areas of research. Students can use the same platform to
   exchange their own examples, make suggestions and point out interesting
   concepts. In such a model, the possible contribution from others to the
   training material needs to be moderated and validated with appropriate
   policies.
   \item If complemented by the availability of remote virtual machines
   (possibly via a browser), students could have access to examples and
   exercises that are already embedded in their own natural environment: all the
   necessary tools and libraries needed to implement the exercise will be
   already available in the virtual machine (possibly a Docker container). With just a
   web browser, students could run complex examples from home, taking advantage
   of a remote facility that provides some storage and computing power.
   Important here is the concept of ``environment'': a Docker container could be
   set up in such a way that students will work in an exact replica of the
   environment they will be exposed to in their experiment. Moreover, students
   could be provided with Docker containers that
   preserve their modified environment across sessions, allowing them to
   develop their skills over a prolonged period of time by accessing all the
   files that were made persistent day by day during the training.
   \item There would no longer be a need to find the resources to host a school
   and pay the tutor(s) (and eventually subsidize the students to participate in
   training in a remote location). Students could follow the training at their
   own pace from wherever they happen to be. A traditional school only lasts for
   5 days (10 at most) and it is difficult to cover a  subject to any
   significant depth in such a short time. The web approach, instead, would
   allow for very long and in-depth coverage of any kind of subject, and in this
   sense it could be a {\it complementary} approach to a traditional school. Of
   particular interest could be courses such as ``Machine Learning'',
   ``Statistical Analyis with ROOT'', or even just ``Good practices in C++'' or
   ``Python Programming for scientific computing''.
   \item Finally, this approach could allow the creation of {\it browsable}
   repositories of all training materials, grouped by argument, by relevance, by
   experiment or whatever other critieria. Students from all over the world
   could be exposed to a large repository of examples, exercises and in general
   training material from their own home.
\end{itemize}

An example of a such a web-based platform already exists, and has been
implemented as an Open Source project (backed by Wikimedia) by a group of more
than 30 Italian students. The project is WikiToLearn~\cite{WikiToLearn}. It hosts training material in
several languages, for several disciplines, ranging from Economics to Physics,
Mathematics, and several others. Because it is based upon
wikimedia~\cite{WikiMedia} software, it is very easy
to add material to the site, and to make it appear under a specific topic (such
as Software/Techniques/Machine-Learning) and to manipulate it as if it
were a single document. In the end, students can selectively choose individual
chapters from the site and have the corresponding pdf sent them as a book,
complete with index, content, and chapters.

The adoption of such an approach is made rather easy in WikiToLearn by the
relative simplicity of the wikimedia-based toolset: users contribute their
training material using just a web-browser, and in order to do this efficiently,
the necessary learning curve has been kept appropriately shallow.
An interesting exercise in this context has recently been made by colleagues of
the GridKa School of computing: the material from this year (2017) has been made
publicly available on WTL~\cite{WikiToLearnGridka}.
This constitutes an interesting example of what can be accomplished using this
platform; it is just a first example of what is possible, but an inspiring one.

Another example of web-based platform is a collection of online tutorials
(mostly written in French) hosted on the Gitlab IN2P3
server~\cite{Code-Swim-Coaches}. Taking advantage of
the Gitlab ability to host Docker images, those tutorials aim to avoid the
``damned installation step'' that often absorbs half of a training session.
Similarly to the GridKa School for WikiToLearn, the annual IN2P3 Computing
Days are an opportunity to refresh and extend the collection of
tutorials every year. Future work will focus on English translation, and the
development of a web site which will index the above tutorials, together with
the best recommended external tutorials.

Finally it should be important to evaluate, if and to which extent, the
complementary approaches to training, such as schools and dedicated web-site
courses, could be of mutual benefit, in other words how to make them efficiently cooperate in the
development of a complete training program.

\subsection{Enhancing Current Training Programmes}
\label{sec:ehcurtrainprog}

To achieve our goals for training the community, we can take advantage of
existing training forums. Resources such as conferences, workshops, and schools
(in person and online) can provide a lot of value for our training purposes with
little effort to set up. We should leverage the existing training forums that
most closely match the HEP community's needs.

Within the HEP community, there are already some working examples of dedicated
training environments that alternate between general topics and
experiment-specific topics. The LHC Physics Center (LPC) at Fermilab hosts
Hands-on Advanced Tutorial Sessions
(HATS)~\cite{LPCHandsOn}
throughout the year to introduce and train participants in topics as diverse as
the latest $b$-tagging algorithms, Git/GitHub, and machine learning. These HATS
provide face-to-face time with instructors and participants at Fermilab, and
also allow remote collaborators to join in by video and complete the same online
exercises. A similar approach is in use in the CMS Data Analysis Schools
(CMSDAS)~\cite{LPCDataAnalysis},
a series of week-long workshops that now take place at multiple labs all over
the world and are designed to ramp up new collaborators in CMS-specific analysis
tools while providing some discussion of the physics as well.

Other examples are CERN School of Computing~\cite{CERNSchoolOfComputing},
CERN OpenLab Software workshops~\cite{OPENLab} education in collaboration with
industry partners, and a series of more advanced topical training courses
provided by MPI Munich and DESY that focus on advanced programming, use of
acceleration hardware and statistical tools including machine learning. This
list includes the
Bertinoro~\cite{Bertinoro},
GridKa~\cite{GridKa}, and
CODAS-HEP~\cite{CODAS-HEP} Schools of Computing.

Over the past decade, MOOCs have been developed by universities and private organizations.
They have been well received by industry and academia.
In addition, they provide a lot of flexibility in terms of cost and use of time;
they are typically free and open for enrollment at any time of the
year. Since the material can be accessed at any time and revisited at any time,
they can be completed at a pace that makes sense, for example, for a physicist
who needs to learn machine leanrning in a piecemeal way.

There are a growing set of MOOCs teaching various subjects. Since there are many
options, there is a wide variety with respect to the depth of the material and
specific tools taught. Exploring these options allows us to choose which is the
right offering for the knowledge needed to work on a specific experiment. We
can pick and choose modules to tailor an appropriate roadmap of skills to learn.
More difficult will be to assemble specific training material not already
avaible elsewhere in an efficient and organized way, since this requires
adequate organization, volunteers and a suitable infrastructure.

Several industry conferences already exist that bring together those in academia
and industry who are at the cutting edge of these techniques. Conferences such
as NIPS~\cite{NIPS} and
PyData~\cite{PyData} provide a focused place where
attendees can learn a lot about machine learning in a short period of time.
Machine learning concepts such as current methods, tools, and problems facing
industry and academia can be learned at conferences. In addition, conferences
are an excellent networking opportunity; attendees can meet and share ideas with
fellow learners and experts. Bonds can be formed quickly at conferences that can
be maintained after the duration of the conference. These connections to the
outside community can be essential since we will be evolving training materials
to ensure that they stay relevant over time.

For example, at the time of this writing,
Coursera~\cite{Coursera} and
Udacity~\cite{Udacity}
both provide great machine learning massive open online courses at no cost.
These two courses both provide a great foundation for assimilating machine learning
fundamentals. However, Coursera's approach emphasizes more theory (with more
math background necessary) and uses MATLAB/Octave while Udacity's approach
emphasizes more practical aspects using machine learning techniques
using Python tools.

\subsection{Resources and Incentives}

It should be considered that some graduate student
advisors, might need to be encouraged to make sure their students are properly
trained. Sometimes, students are instead pushed to learn the bare minimum to get
the work done, at the expense of a broader training/education curriculum that
could actually yield improved results further down the line. One incentive
would be to provide training programmes that also count as course credit, perhaps
as an elective. This model is already in limited use with some online
solutions~\cite{TrainingProgram}, but this is not universal.
Discussions should be started with collaborators at higher education
institutions to see what the roadblocks or opportunities would be for these
training sessions to serve double duty.
It should also be considered that not all students will end their career in
research or academia: their contribution to the research activity, as students,
should therefore also provide them knowledge, know-how, and skills considered a
valuable asset by industry, in order to increase their chances of a career
outside research.

Training is something that a large cross-section of the community
understands to be important, but finding time and effort to contribute to this
project is not actively on the radar of most potential volunteers. Providing
incentives for their participation and creating the appropriate platforms can go
a long way to reach a productive training environment. It can be as simple as
inviting someone to give a software tutorial on the subject that they are
familiar with, give a lecture or seminar or contribute to a growing knowledge
base.

Another important incentive is recognition. For younger members of the
community, having the opportunity to create a training resource, such as a
software tutorial or a knowledge base on a particular topic, is very empowering
and motivational to continue the efforts of training others. Engaging younger
members of the community is crucial to long-term success of HEP training
endeavors.

In the context of web-based training, if the HSF helps to constitute a living
collection of online tutorials, we could organize regular events such as
a ``Tutorial of the month'', or some sort of ``like'' system for the students
to support their favorite tutorials. The best authors must be recognized, so
that they can showcase their most popular online tutorials, just the same as
their research publications.

It is also critical to incorporate training into grant proposals so that it can
be connected with other areas such as research and development. Efforts like
DIANA-HEP~\cite{DIANA-HEP} and AMVA4NewPhysics~\cite{AMVA4NewPhysics}
that combine training
and software development are good examples of such ideas in practice. More
examples of such efforts are needed.

\section{Other resources}

Software Carpentry~\cite{SoftwareCarpentry} and Data
Carpentry~\cite{DataCarpentry} are two parts of The
Carpentries organization that collaboratively build and teach some of the basic
concepts in developing and maintaining software, and analyzing data,
respectively.  The materials that they develop and use are open, and can be
customized for science domains (e.g., HEP), or participant groups (e.g.,
undergraduates). Their model is that they offer training of trainers, and then
the trainers who have graduated can offer training under the SC/DC names, though
of course, anyone can use the SC/DC materials without doing it under the SC/DC
names.
Subjects covered in a typical Data Carpentry school are given in Appendix
\ref{appendix-b}.

While Software Carpentry is leveraged to build a foundation of knowledge for
later more advanced concepts, it is important to note much of this material is
developed specifically for this course and not a part of a larger Software
Carpentry programme. This course focuses on a shallow but wide building of
foundational skills approach, introducing many base concepts but not going deep
into any one concept, with a through line of Open Science and Ethical Data
Usage.

\section{Conclusions}

The HEP community by and large acknowledges and recognizes the great importance
of training in the field of scientific computing. This activity should encompass
several types of \emph{students}, from undergraduates, to young researchers,
up to senior physicists, all of them in need of an appropriately designed training
path in order to be proficent in their scientific endeavours.

We have identified a certain number of problems that need to be overcome in
setting up an appropriate training programme:
\begin{itemize}
    \item Costs and relative funding
    \item Incentives
    \item Career paths
    \item Overall organization across experiments, countries and
    corresponding Funding Agencies
\end{itemize}

For each of these points we have provided an overview of the current situation
and made proposals to improve the situation in HEP. The ideas presented need to
be developed further and concrete actions in the community need to be
implemented, which we will undertake in the HSF Training Working Group\cite{HSFTrainingWG}.

\newpage
\begin{appendices}

\sloppy
\raggedright

\hypertarget{appendix-a---training-topics-of-interest}{%
\section{Training Topics of Interest}\label{appendix-a---training-topics-of-interest}}

\begin{itemize}
   \item Basic and Advanced Programming Concepts
   \begin{itemize}
      \item Object oriented paradigm
      \item Compiled langauges (C++)
      \item Scripting languages (Python, Javascript,...)
   \end{itemize}
   \item Algorithms
   \begin{itemize}
      \item Boost library
      \item STL algorithms for containers
      \item R and/or ROOT
   \end{itemize}
   \item Frameworks (development or application level)
   \begin{itemize}
      \item Qt
      \item ROOT
      \item experiment specific frameworks (possibly if of potential interest outside the originating experiment)
   \end{itemize}
   \item Code design (design patterns)
   \item Development tools
   \begin{itemize}
      \item IDEs (Integrated Development Environment)
      \item Debuggers
      \item Profilers
   \end{itemize}
   \item Evaluation metrics
   \item “Trust” metrics such as data driven tests
   \item Specific software implementation training
   \item Good practices
   \item Code style and clarity
   \item Scripting and data cleaning
   \item Reporting results reproducibly
   \item Writing Documentation
\end{itemize}

\hypertarget{appendix-b}{%
\section{Research Data Science Curriculum}\label{appendix-b}}

This list is taken from the curriculum of the CODATA-RDA Research Data
Science Summer School in progress in Trieste, Italy during July 2017.
(http://indico.ictp.it/event/7974/):

\begin{itemize}
\item Introduction
\item UNIX Shell programming (Software Carpentry Module)
\item GitHub (Software Carpentry Module)
\item R (Software Carpentry Module)
\item BYOD (Bring Your Own Data) best practices
\item Data Management with SQL  (Software Carpentry Module)
\item RDM Storage Management
\item Visualisation
\item Machine Learning Overview - Recommendation
\item Recommender Systems
\item Artificial Neural Networks and other Machine Learning Systems
\item Research Computational Infrastructure
\end{itemize}

\end{appendices}

\sloppy
\raggedright
\clearpage
\printbibliography[title={References},heading=bibintoc]

\end{document}